\documentclass[twocolumn,showpacs,preprintnumbers,amsmath,amssymb]{revtex4}

\usepackage{graphicx}
\usepackage{dcolumn}
\usepackage{bm}
\usepackage{color}
\begin{document}

\title{Exploring the centrality dependence of elliptic and triangular flows}

\author{ V.L.~Korotkikh$^{1}$ and A.M.~Snigirev$^{1,2}$}



\affiliation{$^{1}$ Skobeltsyn Institute of Nuclear Physics, Lomonosov Moscow State University, 119991, Moscow, Russia }

\affiliation{$^{2}$ Bogoliubov Laboratory of Theoretical Physics, JINR, 141980, Dubna, Russia }






\date{\today}
\begin{abstract}
A detailed study of elliptical and triangular eccentricities in the initial state of relativistic  heavy ion collisions is presented. A model of randomly distributed  sources of energy density in the transverse plane based on the effective theory of Color Glass Condensate is used. This model describes well the ALICE and ATLAS data for Pb+Pb collisions at center-of-mass energy 5.02 TeV per nucleon pair in a wide range of centralities, if the second $v_2$ and third $v_3$ harmonics of the anisotropic flow are simply implied to be proportional to the eccentricities $\varepsilon_2$ and $\varepsilon_3$ as a reasonable approximation. The eccentricity  $\varepsilon_2$ is closely related with  the collision geometry and its centrality dependence is mainly determined by the edge diffuseness of the region of the uniform distribution of the saturation pulse of the oncoming nucleus. The eccentricity  $\varepsilon_3$ is completely determined by the chaotic fluctuations of the source position  in the region of overlapping nuclei and is substantially dependent on the overlap area only.
\end{abstract}


\maketitle

\section{Introduction}

Having the LHC put into operation, one got access to a number of experimental intriguing and exquisite phenomena which would have never been systematically studied at the accelerators of previous generations. In this paper we explore and draw attention to the centrality dependence of elliptic and triangular flows as a power probe of collective properties of sub-nuclear matter created in relativistic heavy ion collisions (see, e.g., recent Proceedings of Quark Matter 2018~\cite{Proceedings:2019drx}). Such dependence has been thoroughly measured by the ALICE~\cite{ALICE} and ATLAS~\cite{ATLAS_2019} Collaborations in Pb+Pb collisions at center-of-mass energy 5.02 TeV per nucleon pair. In particular, these measurements demonstrate the nontrivial centrality dependence of the ratio of the second flow harmonic to the third one, which is typically missed in hydrodynamical calculations~\cite{Alba:2017hhe} and in many phenomenological approaches (for instance, in the popular HYDJET++ model~\cite{Lokhtin:2008xi,Bravina:2013xla,Bravina:2013ora}). A simple model of event-by-event fluctuations of energy density proposed in Ref.~\cite{Gelis} reproduces naturally the experimental data for the ratio above for the central collisions (up to $30\%$)  and here is applied to investigate the centrality dependence of elliptic and triangular flows in detail.

Our article is organized as follows. The basic principles of the approach are given in Sec.~II. The numerical results are compared with the experimental data in Sec.~III. Conclusions are drawn in Sec.~IV.

\section{Theoretical framework}

The resulting  profile of energy density in an ultrarelativistic nucleus-nucleus collision is simulated~\cite{Gelis} as the sum of contributions of elementary collisions between a localized  color charge and a dense nucleus. Each elementary collision yields  a sourse of energy density which is independent of rapidity and decreases  with distance from the center of the source. Thus, the energy density $\rho({\bf r,b})$ as a function of the transverse distance ${\bf r}$  and the impact parameter ${\bf b}$ is determined by the product of the saturation momentum squared of one nucleus and the random source depositions of other nucleus~\cite{Gelis}
\begin{eqnarray}
\rho({\bf r,b}) &=& \sum_{j\subset A} Q^2_B({\bf s}_{A,j},{\bf b})\Delta_A ({\bf r}- {\bf s}_{A,j},{\bf b}) \nonumber\\
&+& \sum_{j\subset B} Q^2_A({\bf s}_{B,j},{\bf b})\Delta_B ({\bf r}- {\bf s}_{B,j},{\bf b}).
\label{eq:Tp7}
\end{eqnarray}
Here $Q_A$ and $Q_B$ are the saturation momenta of the colliding nuclei $A$ and $B$ to be specified below. The positions ${\bf s}_{A,j}$ and ${\bf s}_{B,j}$ are assumed to be independent random variables. 

The profile $\Delta$ of energy source in nucleus ($A/B$) is selected in the form which is satisfied the short distance correlations in the Color Glass Condensate (CGC) approach~\cite{cgc1,cgc2,cgc3}:
\begin{eqnarray}
& \Delta ({\bf r}- {\bf s}_{A,j},{\bf b}) \nonumber \\
& =  \left\{ \begin{array}{ll}
\frac{8}{g^2 N_c}\frac{1}{|{\bf r} - {\bf s}_{A,j}|^2 +Q^{-2}_A( {\bf r},{\bf b}) }, & |{\bf r}- {\bf s}_{A,j}|<1/m\\
0, & |{\bf r}- {\bf s}_{A,j}|>1/m
\end{array} \right\},
\label{eq:Tp5}
\end{eqnarray}
where $g$ is the dimensionless coupling constant of QCD, $N_c$ is the number of colors ($N_c=3$ for QCD) and $m$ is the infrared cutoff parameter of the order of the pion mass. At the large distance  $\Delta{\bf (r)}$ decreases like $1/r^2$ as  for a Coulomb field in two dimensions. However,  $\Delta{\bf (r)}$ goes to a finite value for $r \rightarrow 0$, while it would diverge for a pointlike charge. The physical interpretation is that the charge is spread over a distance $\sim 1/Q_A$. The number of elementary charges contained in an area of this size is of order $1/g^2$, which explains the normalization factor in Eq.~(\ref{eq:Tp5}).

If a source is located  in the region of the nuclear size  then the distribution~(\ref{eq:Tp5}) is concentrated inside an area with a radius  $|{\bf r}| <1/m = 1.4$ fm.  It is  considerably smaller than the transverse area of heavy  nuclei (for instance,  for Pb with the radius $ R=6.62$ fm). At  $ {\bf s}_{A,j} = {\bf r}  $ the energy intensity  is maximum in the nuclear center and is proportional to
$Q^{2}_A( {\bf r,b}) $. The integral intensity of one source is equal to 
\begin{eqnarray}
I_A({\bf r,b}) &=& \int d^2s \Delta ({\bf r}- {\bf s},{\bf b})\nonumber \\
&=& \frac{8 \pi}{g^2N_c} 
\ln(1+\frac{Q^{2}_A( {\bf r,b})}{m^2 }).
\label{eq:Tp6}
\end{eqnarray}

In the so-called magma model~\cite{Gelis} $Q^2_A$ is assumed to be proportional to the integral of the nuclear  density over the longitudinal coordinate $z$, i. e. to the thickness function:
\begin{equation}
Q^2_A(x,y) = Q^2_{s0}T_A(x,y)/T_A(0,0),
\label{eq:Tp1}
\end{equation}
where this thickness function is defined as
\begin{equation}
T_A(x,y) = A \int dz  \; \rho_A(x,y,z)
\label{eq:Tp2}
\end{equation}
and  the three-dimension nuclear density $\rho_A(x,y,z)$  is determined by the standard Fermi-Dirac (or Woods-Saxon) distribution 
\begin{equation}
\rho_A(x,y,z)=\rho_0 \frac{1}{ e^{(r-R)/ d} +1}.
\label{eq:Tp3}
\end{equation}
Here $R$  is the nuclear radius, $A$ is its atomic number, $d$ is the diffuseness edge parameter  and $\rho_0$ is a normalization constant so that $\int d^3r  \; \rho_A(r)=1$.  The value of the saturation momentum $Q_{s0}$ at the nucleus center is a free parameter in this approach.

The number of sources per unit area (density) is determined in accordance with the distribution (\ref{eq:Tp1}) and is equal to
\begin{equation}
n_A  = \frac{N_c^2}{32\pi}\frac{Q^2_A({\bf r,b})}{\ln(1+\frac{Q^{2}_A( {\bf r,b})}{m^2 })}.
\label{eq:Tp8}
\end{equation}
Thus, the maximum number of sources in the nuclear transverse area with  $R=6.62$ fm 
at  $Q_{s0} = 1.24$~GeV is estimated as $N_ {\rm PB} = 96$. This number characterizes the fluctuation scale. 

The needed eccentricities are calculated by the standard formulas(for simplicity we omit the variable {\it b} ):
\begin{eqnarray}
\varepsilon_n &=& \varepsilon_{n,x} + i\varepsilon_{n,y} = 
\frac{\int s d s d\phi e^{i n\phi} s^n\rho ({\bf s}) }
{\int  s d s d\phi s^n\rho({\bf s})},\nonumber\\
|\varepsilon_n|^2 &=&  \varepsilon_{n,x}^2 + \varepsilon_{n,y}^2,\nonumber \\
\varepsilon_n\{2\} &=& \sqrt{< |\varepsilon_n|^2 >},
\label{eq:Tp9}
\end{eqnarray}
where angular brackets denote an average value over many events in a narrow centrality  class.
The effective transverse overlap area between the two nuclei is defined as~\cite{dd2010}
\begin{equation}
S(b) = 4\pi \sqrt{<x^2>}\sqrt{<y^2>}.
\label{eq:Tp10}
\end{equation}
We note that there is no commonly accepted definition of the absolute normalization of the overlap area. Our area definition with maximum magnitude $4\pi$ is four time larger than that defined, for instance, in Ref.~\cite{voloshin} but coincides practically with the geometrical overlap area of two disks with uniform two-dimensional distribution of density.

\section{Results}
The formalism briefly reviewed in the previous Sec.~II is applied to calculate the initial eccentricities $\varepsilon_2$ and  $\varepsilon_3$ as functions of the geometrical centrality $C=b^2/(4R^2)$ in relativistic heavy ion collisions. One should note that anisotropic flow is not measured on an event-by-event basis. Values accessible experimentally are moments or cumulants of the distribution of the flow harmonic coefficients $v_n$. 
The lowest order cumulants are defined as~\cite{borghini}:
\begin{eqnarray}
v_n\{2\} &=& \sqrt{< |v_n|^2 >}.
\label{eq:vn}
\end{eqnarray}
In the linear response approximation they are simply proportional to the corresponding cumulants of the initial eccentricities:
\begin{eqnarray}
v_2\{2\} &=& k_2 \varepsilon_2\{2\},\nonumber\\
v_3\{2\} &=& k_3 \varepsilon_3\{2\}.
\label{eq:vn-e2}
\end{eqnarray}

For the central collisions (up to $30 \%$) the magma model describes successfully~\cite{Gelis} 
the experimental data on $v_2$ and $v_3$ as functions of centrality percentile, 
measured by the ATLAS Collaboration~\cite{ATLAS_2019} in 5.02 TeV Pb+Pb collisions. 
The proportionality coefficients $k_2=0.321$ and $k_3 =0.314$ together with 
the saturation momentum $Q_{s0}=1.24$ GeV were adjusted to data. 

We reproduce numerically the results of Ref.~\cite{Gelis} with the same free parameters
$k_2=0.321$, $k_3 =0.314$ and $Q_{s0}=1.24$ GeV fixing the nuclear diffuseness edge 
parameter $ d = 0.1$ fm. At such a small value of $d$ the eccentricity $\varepsilon_2$ 
calculated in the magma model is close to the pure geometrical one, $\varepsilon_{2,\rm geom}=b/(2R)$, 
obtained in the hard sphere model for the centrality interval $ 2-30\%$. 
For the very small centrality $0-2\%$  the magma eccentricity goes to a finite value, 
while the geometrical $\varepsilon_2$ goes to zero at $C \rightarrow 0$. 
For the large centralities ($> 30\%$) a linear dependence $ v_n(C)=k_n \varepsilon_n(C)$ 
with constant coefficients $k_n$ is not realized. However, the hydrodynamic calculations 
indicate that the linear response $ v_2(C)=k_2(C) \varepsilon_2(C)$ is still  possible 
in a wide centrality region if one uses the correct centrality dependence of $k_2(C)$    .

\begin{figure}[htpb]
\label{fig1}
\includegraphics[width=8.00cm]{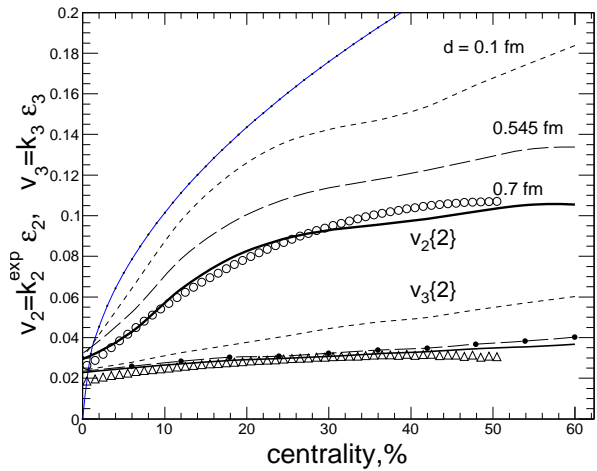}
\caption{ The elliptic and triangular flow cumulants in the magma model. The ALICE data 
in 5.02 TeV Pb+Pb collisions are open icons. The model predictions at the different diffuseness  parameter $d$ are the lines.
The dotted line is the model result at $d = 0 $~fm.   $k_3 = 0.321$ .  }
\end{figure}

In nuclear physics the edge parameter $d=0.545$ fm is commonly used to describe the diffuseness of 
the nucleon density on the nucleus edge. Moreover, the diffuseness of gluon field edge 
in a nucleus can be even larger than the nucleon diffuseness. Therefore, we investigate 
the influence of the edge parameter on the centrality dependence of eccentricities. 
The results are presented in Fig.~1 at the different diffuseness  parameters $d$. 
The best agreement with experimental data is obtained at  $d = 0.7$ fm and $k_2^{exp}$ ~\cite{ALICE}. 
At the sharp nuclear edge the overlap region is strongly pronounced. It is not the same at the large diffuseness. 
The eccentricity $\varepsilon_2$ becomes smaller at the large centralities. This result indicates a significant role of the gluon field edge region for the second harmonic. The third harmonic is weakly sensitive to the diffuseness variation, i.e. to the shape of the overlap region.

Unlike the second harmonic determined by the overlap region shape (the collision geometry) mainly, the third harmonic has a pure fluctuation origin and its magnitude is practically  determined by the overlap region area, but not its shape. The indirect evidence of this affirmation follows from the fact that the ratio  $\sqrt{<v^2_3> - <v_3>^2}/<v_3>$  is constant  at all centralities and is  dependent on the transverse momentum slightly.

Figures 2 and 3 illustrate the statement above. Indeed, in the simple model of hard  sphere we can easily calculate the are of overlap region $S(C)\simeq \pi R^2 (1-\sqrt{C})$ and therefore the number of sources $N_{\rm sources} = {\rm density} \times S(C)$. If the third harmonic has a pure fluctuation origin then its magnitude should be 
\begin{equation}
v_3(C)  \sim \frac{1}{\sqrt{N_{\rm sources}}} \simeq \frac{K_3}{ \sqrt{ (1-\sqrt C ) }  }.
\label{eq:Tp11}
\end{equation}
This pure fluctuation centrality dependence~(\ref{eq:Tp11}) is shown in Fig.~2 in comparison with experimental data. At $K_3=0.0183$ we obtain  a very good simple fit practically in the all centrality region.

Moreover, the area of the overlap region has the significance only, but not its shape,
that is illustrated in Fig.~3.  In this Figure we show the magma triangular flow $v_3$ calculated at $b=0$, but with the radius

\begin{widetext}

\begin{figure}[htpb]
\hspace{-0.5cm}
\mbox{
\includegraphics[width=8.00cm]{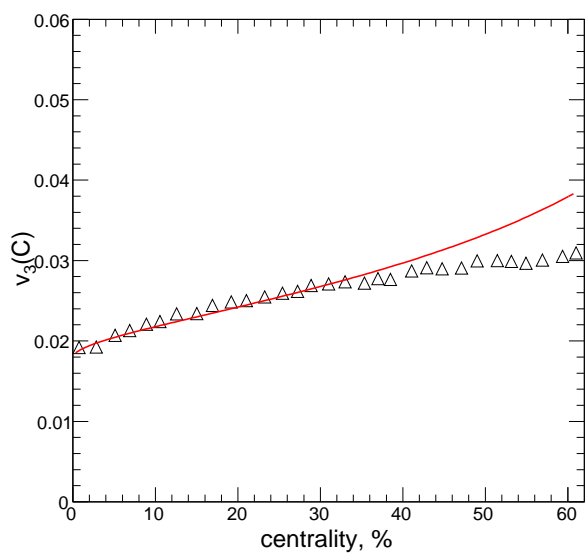} 
\includegraphics[width=8.00cm]{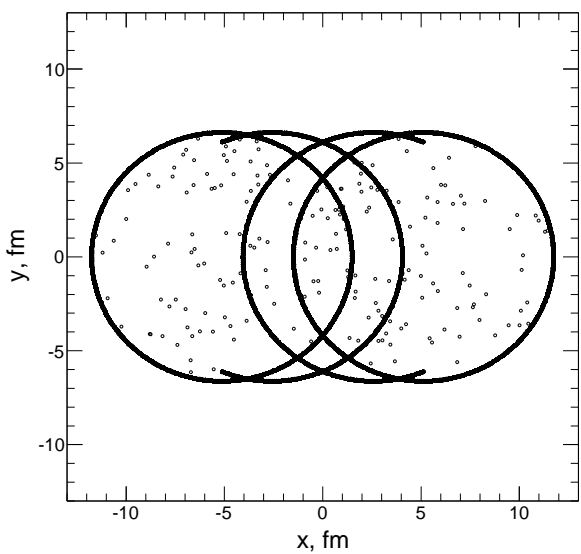}
} 
\caption
{The line is the centrality dependence in accordance with Eq.~(\ref{eq:Tp11}) at $K_3=0.0183$.
Open icons are the ALICE data~\cite{ALICE} (left).
The changing of the overlap region  with the variation of the collision centrality ($C = 60\% $ and $12\%$) is presented for the illustration (right). }
\end{figure}

\begin{figure}[htpb]
\hspace{-0.5cm}
\mbox{
\includegraphics[width=8.00cm]{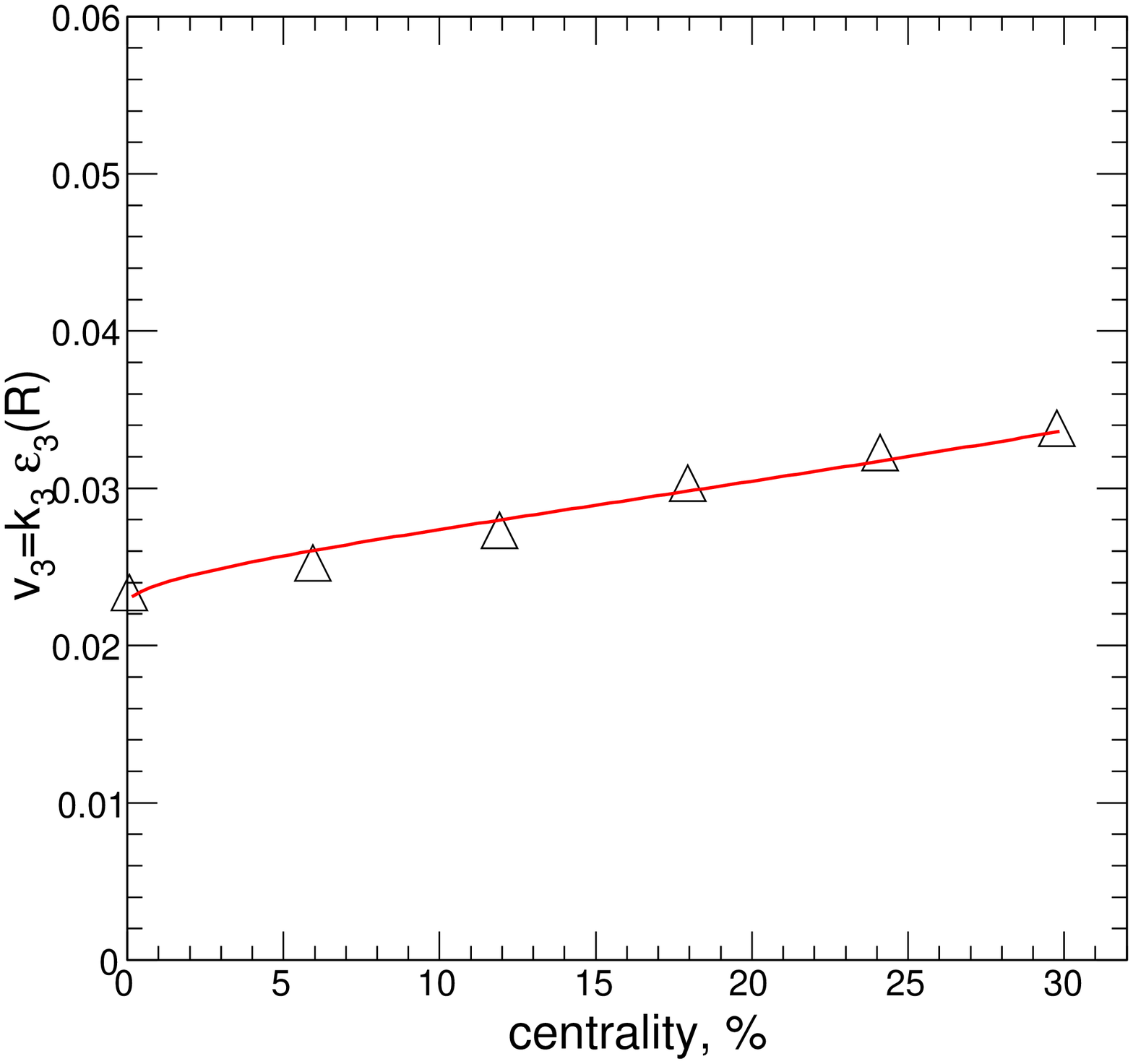} 
\includegraphics[width=8.00cm]{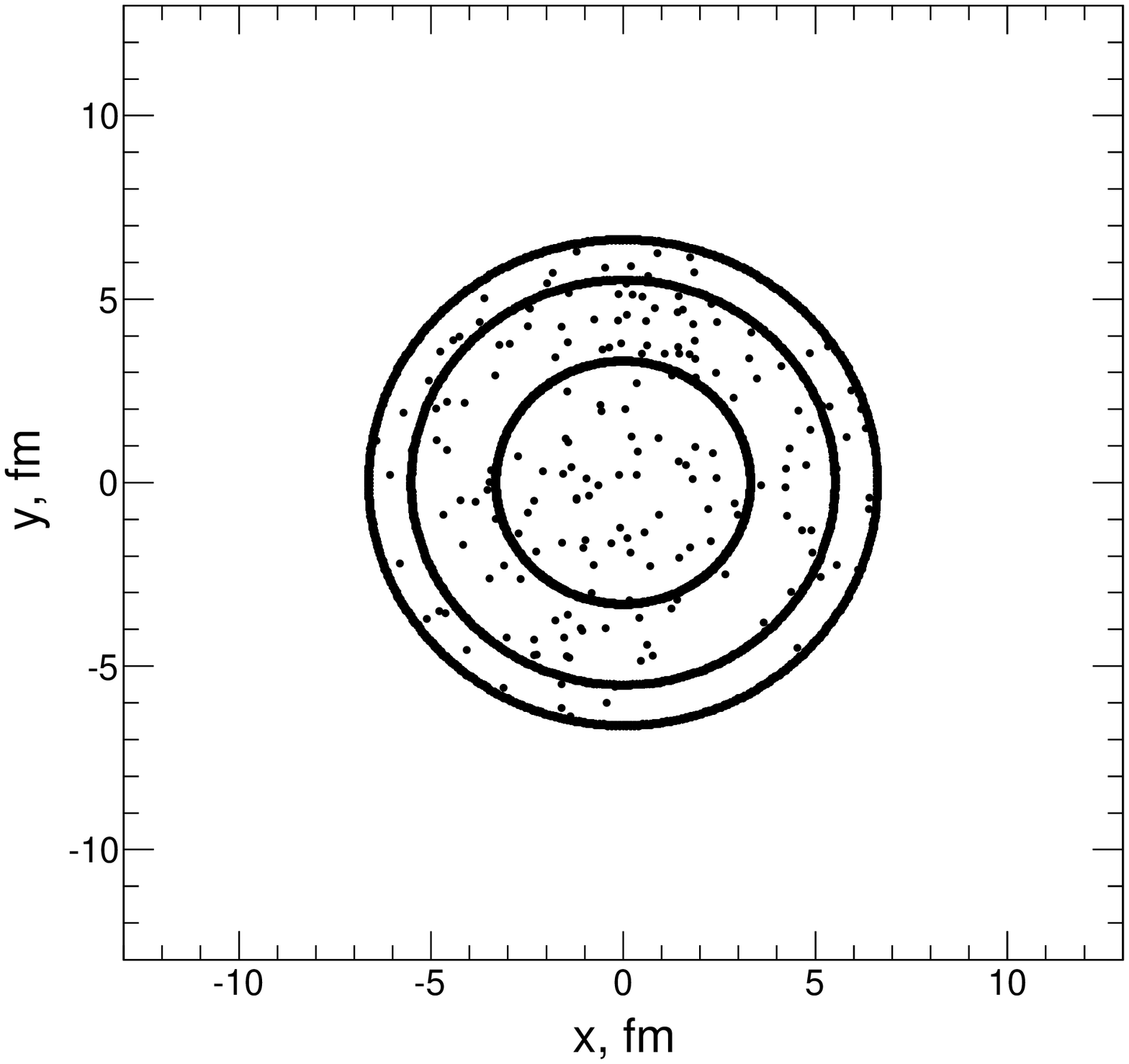}
} 
\caption
{The line is the triangular flow in the magma model at $b=0$ with the radius variation in accordance with  Eq.~(\ref{eq:Tp12}) and at $k_3 = 0.022$ in Eq.~(\ref{eq:vn-e2}).
Open icons are the ALICE data~\cite{ALICE} (left). The changing of the overlap region  with  the radius variation  ($R = 6.62, 5.3, 3.1$ fm) is presented for the illustration (right).}
\end{figure}

\end{widetext}

 \begin{equation}
R (C) = R \sqrt{1 - \sqrt{C}}.
\label{eq:Tp12}
\end{equation}
In this  case the overlap region has a shape of a circle (Fig.~3, the right panel), but with the same area as the overlapping in the hard sphere model at the given centrality $C$. Again the good agreement with experimental data takes place.


The ``pure fluctuation'' contribution to the second harmonic reveals itself in the most central collisions ($0-2\%$) only, where the ``pure geometrical'' contribution goes to zero. This explains also the interesting observation that the ratio $v_2/v_3=1$ at $C=0$. One should note that the absolute magnitude both harmonics in this centrality interval is determined by the source number, i.e. by the area of overlapping at last. It means that magma model predicts the larger value of both harmonics for the lighter nuclei:
\begin{equation}
v_n(C=0)  \sim \frac{1}{\sqrt{N_{\rm sources}}} \sim A^{-1/3}.
\label{eq:v-n,A}
\end{equation}

\section{Conclusion}
 
Our investigation shows that the third harmonic has a pure fluctuation origin and its centrality dependence is determined by the variation of the overlapping area with the changing of the centrality $C$ and is well fitted by the simple ``fluctuation'' formula~(\ref{eq:Tp11}). The elliptic flow  coefficient  $v_2$ is closely related with  the collision geometry, and its centrality dependence is mainly determined by the edge diffuseness. The ``pure fluctuation'' contribution to the second harmonic reveals itself in the most central collisions ($0-2\%$) only. In this centrality interval
the absolute value of all harmonics is simply determined by the number of sources and is independent of the shape of the overlapping region. The magma model gives the interesting prediction for the behavior of magnitude of all harmonics with the variation of atomic number $A$ (scaled as $A^{-1/3}$) in the most central collisions ($0-2\%$).

The calculated initial eccentricities $\varepsilon_2 $ and $\varepsilon_3 $ can be used as an input to the phenomenological models like the HYDJET++ to improve the description of the centrality dependence of the flow azimuthal characteristics.

\begin{acknowledgments}
Discussions with A.I.~Demyanov and I.P.~Lokhtin are gratefully acknowledged. The paper was partially supported by Russian Foundation for Basic Research (grant 18-02-00155).
\end{acknowledgments}


\end{document}